\documentclass[twocolumn,aps,prl,superscriptaddress,showpacs,floatfix]{revtex4-1}
\usepackage[pdfborder=0 0 0]{hyperref}
\usepackage[dvips]{graphicx}
\usepackage{amsmath}
\usepackage{color}
\usepackage{amsfonts}
\usepackage{amssymb}
\usepackage{epstopdf}
\usepackage{colortbl}
\usepackage{nicefrac}
\DeclareGraphicsExtensions{.eps,.png,.pdf}
\begin{document}
\title{An enhanced operating regime for high frequency capacitive discharges}
% with the application of a weak magnetic field}
%\title{Enhancement of a high frequency capacitive discharge at a low transverse magnetic field due to synchronization between electron motion and sheath oscillation}
\author{Sanket Patil}
\affiliation{Institute for Plasma Research, Bhat, Gandhinagar 382 428, India}
\author{Sarveshwar Sharma} 
\email[e-mail: ]{sarvesh@ipr.res.in}
\affiliation{Institute for Plasma Research, Bhat, Gandhinagar 382 428, India}
\affiliation{Homi Bhabha National Institute, Anushaktinagar, Mumbai 400 094, India}
\author{Sudip Sengupta} 
\affiliation{Institute for Plasma Research, Bhat, Gandhinagar 382 428, India}
\affiliation{Homi Bhabha National Institute, Anushaktinagar, Mumbai 400 094, India}
\author{Abhijit Sen}
\affiliation{Institute for Plasma Research, Bhat, Gandhinagar 382 428, India}
\affiliation{Homi Bhabha National Institute, Anushaktinagar, Mumbai 400 094, India}
\author{Igor Kaganovich}
\affiliation{Princeton Plasma Physics Laboratory, Princeton University, Princeton, New Jersey 08543, USA}
\newcommand{\beq}{\begin{equation}}
\newcommand{\eeq}{\end{equation}}
\newcommand{\beqstar}{\[}
\newcommand{\eeqstar}{\]}
\newcommand{\bea}{\begin{eqnarray}}
\newcommand{\eea}{\end{eqnarray}}
\newcommand{\beastar}{\begin{eqnarray*}}
\newcommand{\eeastar}{\end{eqnarray*}}
\begin{abstract} 
We report the existence of an enhanced operating regime for a high-frequency, low-pressure capacitively coupled plasma (CCP) discharge  in the presence of a weak magnetic field applied parallel to the electrodes. Our PIC-MCC simulations show that the plasma density and ion flux values exhibit a sharp peak when the electron cyclotron frequency equals half of the applied RF frequency.  The physical mechanism responsible for this behaviour is traced to a synchronization between the oscillatory motion of the electrode sheath edge and the motion of a set of electrons reflected by this sheath. These electrons gain a substantial amount of energy that causes a concomitant higher ionisation leading to a peak in the ion flux.  Our theoretical findings should be easy to verify experimentally in present day CCP devices and could provide useful guidelines for enhancing the operational  performance of CCP devices in industrial applications. 
\end{abstract}
\keywords{ }

\pacs{52.80.Pi, 52.50.-b, 52.50.Gj, 52.65.Rr}
\maketitle

Capacitively coupled plasma (CCP) discharges are widely used in plasma processing industry for applications like silicon wafer etching and material deposition. 
The processing rates and the quality of the processed wafer primarily depend on the ion flux incident on the substrate located at one of the electrodes and the energy of those ions. A discharge that yields a higher ion flux  is considered to be more desirable.  Several past studies, carried out on single frequency CCP discharges have shown that it is difficult to have independent control of the ion flux and ion energy \cite{Kaganovich_IEEE_34_2006, Kawamura_POP_13_2006, Lieberman_IEEE_16_1998, Godyak_SJP_78_1976, Godyak_JAP_57_1985, Kaganovich_PRL_89_2002, Sharma_POP_21_2014, Sharma_PSST_22_2013, Sharma_POP_20_2013}. Alternate ideas to overcome this limitation  have been explored in the form of dual-frequency CCPs \cite{Goto_IEEE_6_1993, Robiche_JPDAP_36_2003, Kim_POP_10_2003, Turner_PRL_96_2006, Sharma_JPDAP_46_2013, Karkari_APL_93_2008, Boyle_JPDAP_37_2004, Sharma_JPDAP_47_2014} electrical asymmetric effects \cite{Czarnetzki_PSST_20_2011, Heil_JPDAP_41_2008, Bruneau_PSST_23_2014, Bruneau_PRL_114_2015, Schungel_JPDAP_49_2016} and non-sinusoidal, tailored voltage waveform excitations \cite{Economou_JVST_31_2013, Lafleur_PSST_25_2016, Qin_PSST_19_2010, Shin_PSST_20_2011, Sharma_PSST_24_2015}. The concept of using a magnetic field, applied parallel to the electrodes, to enhance the performance of a CCP has also been explored \cite{Lieberman_91, Hutchinson_95, Park_97}.
However, such magnetic fields, of strengths above a few tens of Gauss, have been found to cause nonuniformity in the plasma and to negatively impact the quality of the wafer \cite{Kushner_2003, Barnat_2008, Fan_2013}.
In this letter, we show that a significant enhancement in the performance of low-pressure CCP discharges can be obtained at a low magnetic field strength ($\sim$ 10 G) if the discharge is operated at very high frequencies.
Simulations carried out at magnetic field strengths ranging from 0 G to 107 G for a 60 MHz, 5 mTorr discharge reveal that the variation of plasma density and ion flux with magnetic field is not monotonic but exhibits a sharp peak at a particular strength of the magnetic field (10.7 G). 
The cause of this enhancement is traced to a synchronization between the oscillatory motion of the electrode sheath edge and the motion of a sub-population of electrons reflected by the sheath. This synchronization leads to a significant energy gain by these electrons leading to a greater ionisation of the neutral background and a resultant sharp rise in the ion flux. The enhanced ionisation mechanism is quite distinct from that caused by strong magnetic fields where the electron confinement time is increased by reducing their cross-field transport  \cite{Lieberman_91, Hutchinson_95, Park_97, Turner_PRL_76_2069_1996}. 
The synchronization occurs when the electron cyclotron frequency ($f_{ce}=\frac{eB}{m_e}$, where B, e and $m_e$ are the external magnetic field, electronic charge and mass respectively) is equal to half the applied frequency ($f_{rf}$).
As $f_{ce}$ depends on the magnetic field, the aforementioned peak in plasma density occurs at a specific strength of the magnetic field, which turns out in our case to be quite low ($ \sim $ 10 G).
This is advantageous because traditionally used higher magnetic field strengths (50-100 G) can cause non-uniformity in the plasma.
Our simulations further show such an enhanced operation regime is only  possible for low-pressure (collisionless) discharges operated at very high frequencies and may be the reason why such an effect was missed in past investigations.
For example, Turner et. al. \cite{Turner_PRL_76_2069_1996} conducted a similar simulation study at different magnetic fields for a collisional, low-frequency discharge, but did not observe an increased density at the magnetic field corresponding to the above condition. 
Very high frequency CCP discharges have received significant attention in recent years because of their superior performance compared to traditional low frequency discharges \cite{Wilczek_2015, Sharma_2016, Bera_2005, Sharma_POP_25_080705_2018, Sharma_JPDAP_52_2019, Miller_PSST_15_2006, Upsdhyay_JPDAP_46_2013, Sharma_POP_25_063501_2018, Wilczek_PSST_27_2018, Sharma_POP_26_2019, Sharma_PSST_29_2020}.
Our study reveals a way to use a weak magnetic field to further extend this advantage without compromising on plasma uniformity. 

Our simulations of a plasma bounded by two parallel electrodes have been carried out with the well-tested and widely used 1D-3V Electrostatic Direct Implicit Particle-In-Cell (EDIPIC) code \cite{Sharma_POP_25_080704_2018, Sydorenko_Thesis_2006, Campanell_POP_19_2012, Sheehan_PRL_111_2013, Campanell_APL_103_2013,  Carlsson_PSST_26_2016}. The code is based on the well-established Particle-in-Cell/Monte Carlo Collision (PIC-MCC) method \cite{Birdsall_AH_1991, Hockney_AH_1988}. The types of electron-neutral collisions taken into account are elastic, excitation and ionization collisions. For ions, ion-neutral elastic and charge exchange collisions have been considered. The metastable reactions are not important at low pressures and therefore have been ignored. The cross-sectional data used for the collisions have been taken from well regarded sources \cite{Shahid_JAP_82_1988, Lauro_JPDAP_37_2004}. The code evolves the positions and velocities of electrons and singly ionized ions $Ar^+$. Note that although the code is 1D in position space, it is 3D in velocity space. Thus, it can accurately simulate the ExB motion of charged particles. The neutral gas dynamics is not evolved; the neutral gas is simply distributed uniformly between the electrodes  at a constant temperature of 300 K. Secondary electron emission has also been ignored since at low pressure, it has a negligible effect on the discharge properties \cite{Lieberman_NJ_2005}. The neutral gas used in this study is Argon at 5 mTorr pressure. The frequency and amplitude of the applied voltage are 60 MHz and 100 V, respectively. A sinusoidally varying voltage has been applied between the grounded electrode (GE) and the powered electrode (PE), which are 32 mm apart. Perfectly absorbing boundary conditions have been used, {\it i.e.} all charged particles get absorbed when they interact with the electrodes.

In our simulations the external magnetic field ($B$), applied parallel to the electrodes, has been varied from 0 G to 107 G and the initial electron and ion temperatures have been taken to be 2 eV and 0.026 eV (300 K) respectively. The initial density is $5 \times 10^{15}$ $m^{-3}$. The cell size ($\Delta x$) used is $1/8$ times the initial Debye length ($\lambda_{De}$). The time step ($\Delta t$) is calculated as ${\Delta x}/(\text{maximum expected velocity})$, where maximum expected velocity is four times the initial thermal velocity of electrons. The time step thus satisfies the stability criterion $\omega_{pe} \Delta t < 0.2$, where $\omega_{pe}$ is electron plasma frequency. This time step also resolves the cyclotron motion even at the highest magnetic field used (107 G). The number of super particles per cell is initially 400 giving the total number of particles to be $\sim 5 \times 10^5$.

\begin{figure}
	\centering
	\includegraphics[width=0.5\textwidth, height=0.28\textwidth]{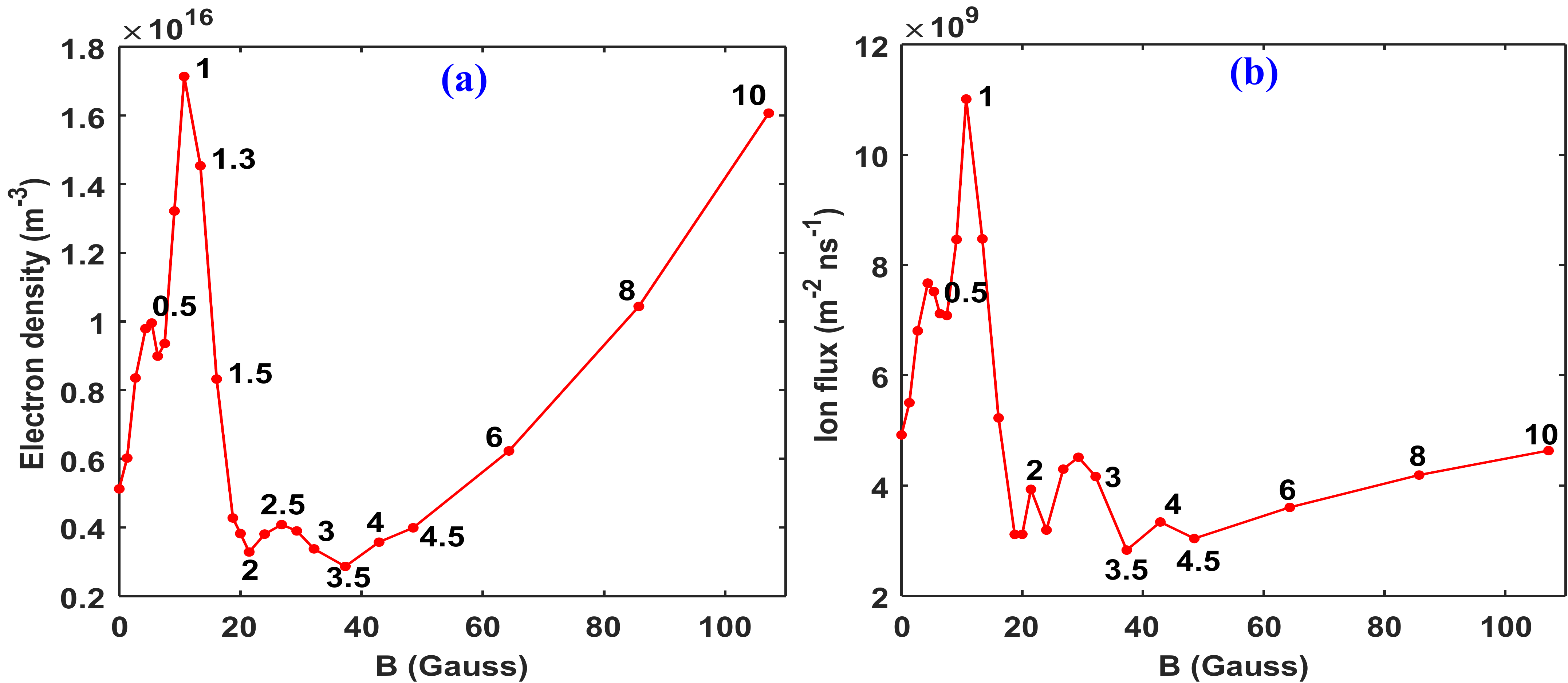}
	\caption{Variation of (a) peak plasma density and (b) ion flux at the grounded electrode with applied magnetic field. The numbers shown against the curves indicate the $r$ value for that particular magnetic field}
	\label{fig:Figure1}
\end{figure}

Our simulation results show that the plasma discharge properties such as the density and the ion flux at the electrodes  do not vary monotonically with the magnetic field  but show significant enhancements at specific strengths of the magnetic field where the ratio of the cyclotron frequency ($f_{ce}$) to the applied RF frequency ($f_{rf}$),
namely, $r = \frac{2f_{ce}}{f_{rf}}$,  assumes simple fractional values like $\frac{1}{4}$, $\frac{1}{2}$, $1$.  Figure ~\ref{fig:Figure1}(a) and Fig.~\ref{fig:Figure1}(b) display this behaviour for the density and ion flux respectively.  The numbers shown against the curves indicate the $r$ value at that particular magnetic field.
\begin{figure}
	\centering
	\includegraphics[width=0.5\textwidth, height=0.28\textwidth]{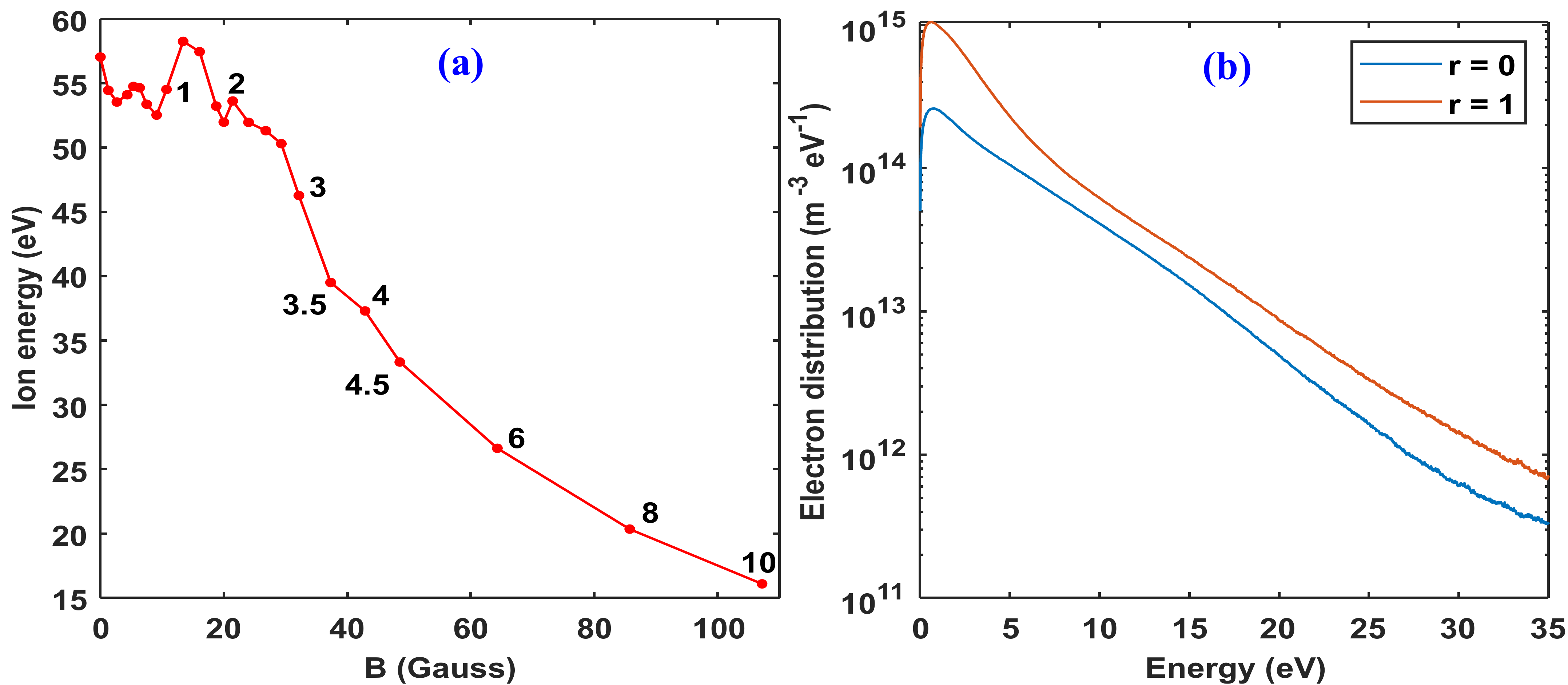}
	\caption{(a) Variation of average energy of the ions incident on the grounded electrode with applied magnetic field (b) Comparison of EEDFs (as measured near the sheath edges) in the absence and presence of the magnetic field. }
	\label{fig:Figure2}
\end{figure}
As can be clearly seen, there is a significant peaking of the plasma density ($1.7 \times 10^{16} m^{-3}$) at $r = 1$ ($B = 10.7$ G) after which it falls and then rises again 
  monotonically above $r = 3.5$ ($B = 37.3$ G) to reach $1.6 \times 10^{16} m^{-3}$ at $r = 10$ ($B = 107.2$ G). The monotonic increase beyond $r=3.5$ can be ascribed to the increase in electron confinement due to the high magnetic field. This is consistent with previous studies on magnetically enhanced CCPs using higher magnetic fields \cite{Sharma_POP_25_080704_2018}. The surprising result is the peak at $r=1$ whose origin we will shortly discuss.
The ion flux ($\Gamma_i = n_iu_i$) at the electrodes, as observed in Fig.~ \ref{fig:Figure1}(b), follows a trend similar to that of the density with a corresponding peak at$ r = 1$. The ion flux however does not increase as much as the density above $r = 3.5$ . This means that at a given amplitude of the applied voltage, the ion flux obtained at $r = 1$ is significantly larger than that in the absence of magnetic field as well as for $ r > 3.5$; large magnetic fields ($r \sim 10$) can produce high discharge densities but not proportionately high ion flux at the electrodes.
To ascertain whether the observed peak in ion flux at $r = 1$ leads to an improvement in the overall performance of the discharge, the average energy of ions incident on the electrodes has to be observed.
Figure \ref{fig:Figure2} (a) shows the variation of the average energy of the ions incident on the grounded electrode with the magnetic field. The average ion energy at $ r = 1$ ($54.5$ eV) is seen to be slightly lower than that in the absence of the magnetic field (57 eV). This change can be ascribed to the decrease in the sheath width due to the magnetic field.  Thus, at $r = 1$, a significantly higher ion flux is obtained at the electrodes with a slightly lower average ion energy. This is an improvement over the unmagnetized discharge.
\begin{figure*}
	\centering
	\includegraphics[width=\linewidth]{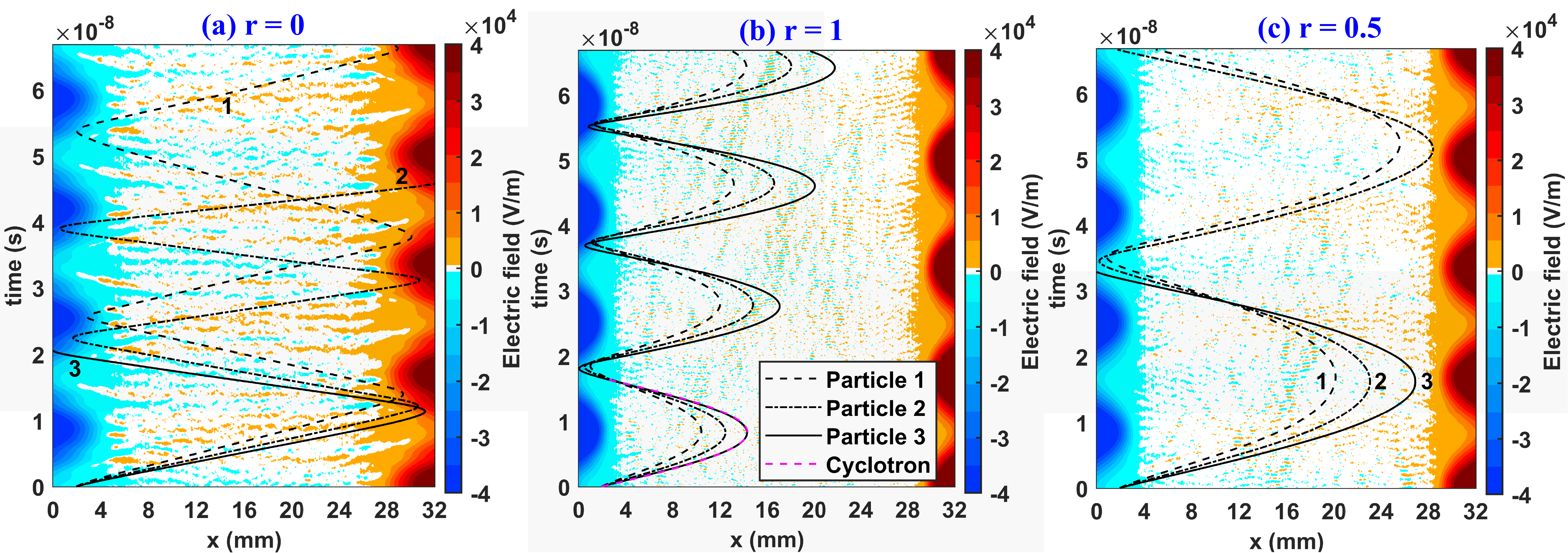}
	\caption{Trajectories of reflected electrons having different energies plotted onto the plot of electric field as a function of space co-ordinate $x$ and time. Particles 1, 2 and 3 have an initial energy of 4 eV, 8 eV and 12 eV respectively. It can be seen that for r = 0, the three particles reach the opposite sheath at different times whereas for r = 0.5 and r = 1, the three particles return to the same sheath simultaneously. For r = 1, the dotted magenta line shows the pure cyclotron motion of a hypothetical particle. It can be seen that the motion of particle 3 closely resembles cyclotron motion.}
	\label{fig:Figure3}
\end{figure*}%
It is clear from Fig. \ref{fig:Figure1}(b) and Fig.~ \ref{fig:Figure2}(a) that operation at $ r > 3.5$, has the advantage of lower ion energies, but at the cost of ion flux. Moreover, using higher magnetic fields has the disadvantage of non-uniformity in the plasma due to {\boldmath$E \times B$} drift, which would be negligible at r = 1. Thus operation at r = 1 is potentially an improvement over the operation at higher magnetic fields (r $\sim$ 10). In Fig.~ \ref{fig:Figure2}(b) we show another distinctive feature of the $r=1$ plasma compared to the $r=0$ case, namely the difference in their respective EEDF. For $r=1$ there exists a higher population of energetic electrons whose role we discuss below in the context of the observed behaviour in Fig. \ref{fig:Figure2}(a) of the density variation  with the magnetic field.\\

We now discuss the physical origin of the surprising peaking of the density for certain values of $r$. For this we examine the trajectories of certain classes of the electron population as they interact with the oscillating sheath edge near the electrode. The potential drop developed in the sheath region is generally much larger than the electron temperatures. Thus, only those bulk electrons that have non-negligible kinetic energies compared to the pre-sheath potential drop are able to reach and interact with the oscillating sheath edges. The trajectories of such electrons, as observed in the simulations, are shown in Fig. \ref{fig:Figure3} for $B = 0 G$, $B= 5.3 G$ (corresponding to $r = 0.5$) and $B=10.7 G$ (corresponding to  ($r = 1$) respectively.  For r = 0 as well as for r = 1, the electrons moving toward an electrode are reflected into the bulk due to the strong electric field in the sheath. It should be noted that  the electron motion, for the $r=1$ case, closely resembles that of cyclotron motion. A comparison between the observed trajectory and the cyclotron orbit is shown in Fig.~ \ref{fig:Figure3} b. This is because when these electrons are reflected from the sheaths, they have such kinetic energies (and hence velocities) that when they enter the bulk plasma the magnetic force on them dominates over  the force due to the weak electric field ($< 10^3 V/m$ in our case, see Fig.~ \ref{fig:Figure3}). Also, the electric field in the bulk oscillates with a high frequency and hence has  very little net effect on the trajectory.
In the unmagnetized case, if a reflected electron does not collide with a neutral in the bulk, it will reach the opposite sheath. Let us refer to the first reflecting sheath as 'sheath 1' (at powered electrode)  and the opposite sheath as 'sheath 2' (at grounded electrode). The relative phase at which this electron arrives at sheath 2 with respect to the phase at which it was reflected by sheath 1 is determined by the time the electron takes to travel the distance between the two sheaths. This travel time is determined by the velocity of the electron after reflection and the electric field it encounters in the bulk, but the latter has a negligible effect as it does not affect the trajectory significantly. Since the electrons reflected by sheath 1 are distributed over a large velocity range, their travel times must also be distributed over a large range. Consequently the relative phases between the electrons arriving at the sheath 2 and that of the RF span a very broad range and do not favour any preferred interaction between them. 
Now consider the trajectory of the reflected electrons for the r = 1 case as shown in Fig.~ \ref{fig:Figure3} b. If an electron does not collide with a neutral, its motion is affected by the magnetic field  and it will return to sheath 1 after completing half a gyration. But as described earlier, the condition r = 1 means that the duration of half a gyration is exactly equal to the duration of an RF cycle. Since the cyclotron frequency ($f_{ce}$) does not depend on velocity, an electron reflected by sheath 1 will return to sheath 1 at the same phase as its reflection, irrespective of its velocity after reflection. Thus, a subset of all the electrons reflected by a sheath at the expanding phase of an RF cycle will return to the same sheath at the same phase and thereby suffer another energy gaining reflection. This process leads to an enhancement of energetic electrons for the $r=1$ case as compared to the $r=0$ case. This is reflected in the EEDFs plotted near to the sheath edges for the two cases as shown in Fig. ~\ref{fig:Figure2}(b). The actual number density of electrons with energies above the ionization threshold  ($15.76$ eV) at the sheath edges are $5.4884\times 10^{13}$ $m^{-3}$ (for r=0) and $9.8364\times 10^{13}$ $m^{-3} $ (for r=1). This excess density of  energetic electrons leads to higher ionization of the neutrals and is the source of the enhanced density peak observed for $r=1$.

A similar enhancement in density and ion flux is also observed for $r = 0.5$. However, the synchronization is less effective in this case for two reasons. Firstly, the reflected electrons return to the sheath after two RF periods, making it more likely (compared to $r = 1$) for them to undergo collisions in the bulk. Secondly, at such a low magnetic field, a significant number of electrons can have a Larmor radius ($r_L=\nicefrac{m_ev_\perp}{eB}$ where $v_\perp$ is the component of velocity perpendicular to the direction of B) large enough to reach the opposite sheath. Synchronization would become impossible for such electrons if their trajectories change significantly upon interaction with the opposite sheath.\\

It is crucial that the discharge be collisionless for the above synchronization process to be effective. Otherwise, most electrons may collide with neutrals in the bulk, and the effects described above will be lost. There is another condition that needs to be satisfied for the synchronization to work, as we will now discuss. We have assumed that an electron reflected by sheath 1 will not interact with sheath 2 during its half gyration, which it can if its Larmor radius is large enough. The distance between the two sheaths for r = 1 in the current study is about 26mm (Fig.~ \ref{fig:Figure3} b) and the cyclotron frequency is 30 MHz. Thus, an electron having Larmor radius greater than $26$ mm must have kinetic energy above $68$ eV. But the population of electrons in this energy range is practically non-existent and can be ignored. This is what makes the synchronization between electron motion and the RF cycle effective. However, if a significant fraction of electrons reflected by sheath 1 interacts with sheath 2, this synchronization would not lead to an increase in density. Consider a low frequency discharge (e.g. $13.56$ MHz) being operated with a small inter-electrode gap, say $32$ mm. At$ r = 1$, $f_{ce} = \frac{f_{rf}}{2} = 6.78 MHz$. Assuming a sheath width of $3$ mm, the Larmor radius of a reflected electron would have to be above $26$ mm to reach the opposite sheath. This Larmor radius corresponds to a kinetic energy of $3.5$ eV at $f_{ce} = 6.78 MHz$. Thus, a significant fraction of electrons would reach the opposite sheath after reflection, which will render the synchronization impossible. If the inter-electrode gap were to be increased to mitigate this issue, the discharge would start to become collisional, and the synchronization will again be ineffective. For example, in the case of \cite{Turner_PRL_76_2069_1996}, a low frequency, collisional discharge had been used, and no notable enhancement at $r = 1$ was observed. Thus, the synchronization is only effective for low-pressure, high-frequency, collisionless discharges.

In conclusion, the combination of a low magnetic field and a high RF frequency could provide an enhanced operational regime for a low pressure CCP device. The physical origin of the enhanced ion flux in such an operational regime appears to be related to a synchronization condition $r = 1$ ($f_{ce} = \frac{f_{rf}}{2}$) which allows a subset of the electron population to gain significant energy from the oscillating sheaths at the electrodes and cause a higher ionization of the neutrals. Our theoretical findings should be easy to verify experimentally in present day CCP devices and could provide useful guidelines for enhancing the operational  performance of CCP devices in industrial applications.

\acknowledgements 
A.S. thanks the Indian National Science Academy (INSA) for their support under the INSA Senior Scientist Fellowship scheme.   I.K.'s research was supported by the U.S. Department of Energy. 

\bibliographystyle{unsrt}

\begin{thebibliography}{10}

\bibitem{Kaganovich_IEEE_34_2006} I. D. Kaganovich, O. V. Polomarov, and C. E. Theodosiou, 
\newblock {\em IEEE Trans. Plasma Sci.} 34, 696 (2006). 

\bibitem{Kawamura_POP_13_2006} E. Kawamura, M. A. Lieberman, and A. J. Lichtenberg, 
\newblock {\em Phys. Plasmas} 13, 053506 (2006). 

\bibitem{Lieberman_IEEE_16_1998} M. A. Lieberman, 
\newblock {\em IEEE Trans. Plasma Sci.} 16, 638 (1988). 

\bibitem{Godyak_SJP_78_1976} V. A. Godyak, 
\newblock {\em Sov. J. Plasma Phys.} 2, 78 (1976).

\bibitem{Godyak_JAP_57_1985} O. A. Popov and V. A. Godyak, 
\newblock {\em J. Appl. Phys.} 57, 53 (1985). 

\bibitem{Kaganovich_PRL_89_2002} I. D. Kaganovich, 
\newblock {\em Phys. Rev. Lett.} 89, 265006 (2002). 

\bibitem{Sharma_POP_21_2014} S. Sharma, S. K. Mishra, and P. K. Kaw, 
\newblock {\em Phys. Plasmas} 21, 073511 (2014). 

\bibitem{Sharma_PSST_22_2013} S. Sharma and M. M. Turner, 
\newblock {\em Plasma Sources Sci. Technol.} 22, 035014 (2013). 

\bibitem{Sharma_POP_20_2013} S. Sharma, and M. M. Turner, 
\newblock {\em Phys. Plasmas} 20(7), 073507 (2013)

\bibitem{Goto_IEEE_6_1993} H. H. Goto, H. D. Lowe, and T. Ohmi, 
\newblock {\em IEEE Trans. Semicond. Manuf.} 6, 58 (1993). 

\bibitem{Robiche_JPDAP_36_2003} J. Robiche, P. C. Boyle, M. M. Turner, and A. R. Ellingboe, 
\newblock {\em J. Phys. D: Appl. Phys.} 36, 1810 (2003). 

\bibitem{Kim_POP_10_2003} H. C. Kim, J. K. Lee, and J. W. Shon, 
\newblock {\em Phys. Plasmas} 10, 4545 (2003).
 
\bibitem{Turner_PRL_96_2006} M. M. Turner, and P. Chabert, 
\newblock {\em Phys. Rev. Lett.} 96, 205001 (2006).

\bibitem{Sharma_JPDAP_46_2013} S. Sharma and M. M. Turner, 
\newblock {\em J. Phys. D: Appl. Phys.} 46, 285203 (2013). 

\bibitem{Karkari_APL_93_2008} S. K. Karkari, A. R. Ellingboe, and C. Gaman, 
\newblock {\em Appl. Phys. Lett.} 93, 071501 (2008). 

\bibitem{Boyle_JPDAP_37_2004} P. C. Boyle, A. R. Ellingboe, and M. M. Turner, 
\newblock {\em J. Phys. D: Appl. Phys.} 37, 697 (2004). 

\bibitem{Sharma_JPDAP_47_2014} S. Sharma and M. M. Turner, 
\newblock {\em J. Phys. D: Appl. Phys.} 47 (28), 285201 (2014)

\bibitem{Czarnetzki_PSST_20_2011} U. Czarnetzki, J. Schulze, E. Schungel, and Z. Donko, 
\newblock {\em Plasma Sources Sci. Technol.} 20, 024010 (2011).

\bibitem{Heil_JPDAP_41_2008} B. G. Heil, U. Czarnetzki, R. P. Brinkmann, and T. Mussenbrock, 
\newblock {\em J. Phys. D: Appl. Phys.} 41, 165202 (2008).

\bibitem{Bruneau_PSST_23_2014} B. Bruneau, T. Novikova, T. Lafleur, J. P. Booth, and E. V. Johnson, 
\newblock {\em Plasma Sources Sci. Technol.} 23, 065010 (2014).

\bibitem{Bruneau_PRL_114_2015} B. Bruneau, T. Gans, D. O’Connell, A. Greb, E. Johnson, and J.-P. Booth, 
\newblock {\em Phys. Rev. Lett.} 114, 125002 (2015). 

\bibitem{Schungel_JPDAP_49_2016} E. Schungel, I. Korolov, B. Bruneau, A. Derzsi, E. Johnson, D. O’Connell, T. Gans, J. P. Booth, Z. Donko, and J. Schulze, 
\newblock {\em J. Phys. D: Appl. Phys.} 49, 265203 (2016). 

\bibitem{Economou_JVST_31_2013} D. J. Economou, 
\newblock {\em J. Vac. Sci. Technol.} A 31, 050823 (2013).

\bibitem{Lafleur_PSST_25_2016} T. Lafleur, 
\newblock {\em Plasma Sources Sci. Technol.} 25, 013001 (2016).

\bibitem{Qin_PSST_19_2010} X. V. Qin, Y. H. Ting, and A. E. Wendt, 
\newblock {\em Plasma Sources Sci. Technol.} 19, 065014  (2010).

\bibitem{Shin_PSST_20_2011} H. Shin, W. Zhu, L. Xu, V. M. Donnelly, and D. J. Economou, 
\newblock {\em Plasma Sources Sci. Technol.} 20, 055001 (2011).

\bibitem{Sharma_PSST_24_2015} S. Sharma, S. K, Mishra, P. K. Kaw, A. Das, N. Sirse and M. M. Turner, 
\newblock {\em Plasma Sources Sci. Technol.} 24, 025037 (2015).

\bibitem{Lieberman_91}
M.~A. Lieberman, A.~J. Lichtenberg, and S.~E. Savas
\newblock {\em IEEE Transactions on Plasma Science} 19(2), 189 (1991).

\bibitem{Hutchinson_95} D.~A.~W. Hutchinson, M.~M. Turner, R.~A. Doyle, and M.~B. Hopkins, 
\newblock {\em IEEE Trans. Plasma Sci.} 23(4), 636 (1995).

\bibitem{Park_97} J.-C. Park and B. Kang, 
\newblock {\em IEEE Trans. Plasma Sci.} 25(3), 499 (1997).

\bibitem{Turner_PRL_76_2069_1996} M.~M. Turner, D.~A.~W. Hutchinson, R.~A. Doyle, and M.~B. Hopkins, 
\newblock {\em Phys. Rev. Lett.} 76(12), 2069 (1996).

\bibitem{Kushner_2003} M.~J. Kushner, 
\newblock {\em J. Appl. Phys.} 94(3), 1436 (2003).

\bibitem{Barnat_2008} E.~V. Barnat, P.~A. Miller, and A.~M. Paterson, 
\newblock {\em Plasma Sources Sci. Technol.} 17, 045005 (2008).

\bibitem{Fan_2013} Yu Fan, Ying Zou, Jizhong Sun, Thomas Stirner, and Dezhen Wang, 
\newblock {\em Phys. Plasmas} 20(10), 103507 (2013).

\bibitem{Wilczek_2015} S. Wilczek, J. Trieschmann, J. Schulze, E. Schuengel, R. P. Brinkmann, A. Derzsi, I. Korolov, Z. Donko, and T. Mussenbrock, 
\newblock {\em Plasma Sources Sci. Technol.} 24(2), 024002 (2015).

\bibitem{Sharma_2016} S. Sharma, N. Sirse, P.~K. Kaw, M.~M. Turner, and A.~R. Ellingboe, 
\newblock {\em Phys. Plasmas} 23(11), 110701 (2016).

\bibitem{Bera_2005} K. Bera, D. Hoffman, S. Shannon, G. Delgadino, and Yan Ye, 
\newblock {\em IEEE Transactions on Plasma Science} 33(2), 382 (2005).

\bibitem{Sharma_POP_25_080705_2018} Sarveshwar Sharma, A. Sen, N. Sirse, M.~M. Turner, and A.~R. Ellingboe, 
\newblock {\em Phys. Plasmas} 25, 080705 (2018).

\bibitem{Sharma_JPDAP_52_2019} S. Sharma, N. Sirse, A. Sen, J. S. Wu, and M. M. Turner, 
\newblock {\em J. Phys. D: Appl. Phys.} 52, 365201 (2019). 

\bibitem{Miller_PSST_15_2006} P. A. Miller, E. V. Barnat, G. A. Hebner, P. A. Paterson, and J. P. Holland, 
\newblock {\em Plasma Sources Sci. Technol.} 15, 889–899 (2006). 

\bibitem{Upsdhyay_JPDAP_46_2013} R. R. Upadhyay, I. Sawada, P. L. G. Ventzek, and L. L. Raja, 
\newblock {\em J. Phys. D: Appl. Phys.} 46, 472001 (2013). 

\bibitem{Sharma_POP_25_063501_2018} S. Sharma, N. Sirse, M. M. Turner, and A. R. Ellingboe, 
\newblock {\em Phys. Plasmas} 25, 063501 (2018).

\bibitem{Wilczek_PSST_27_2018} S. Wilczek, J. Trieschmann, J. Schulze, Z. Donko, R. P. Brinkmann, and T. Mussenbrock,
\newblock {\em Plasma Sources Sci. Technol.} 27, 125010 (2018). 

\bibitem{Sharma_POP_26_2019} S. Sharma, N. Sirse, A. Sen, M. M. Turner, and A. R. Ellingboe, 
\newblock {\em Phys. Plasmas} 26, 103508 (2019).

\bibitem{Sharma_PSST_29_2020} S. Sharma, N. Sirse, A. Kuley, and M. M. Turner, 
\newblock {\em Plasma Sources Sci. Technol.}  29, 045003 (2020).
 
\bibitem{Sharma_POP_25_080704_2018} S. Sharma, I.~D. Kaganovich, A.~V. Khrabrov, P. Kaw, and A. Sen, 
\newblock {\em Physics of Plasmas} 25(8), 080704 (2018).

\bibitem{Sydorenko_Thesis_2006} D. Sydorenko, 
\newblock {\em “Particle-in-cell simulations of electron dynamics in low pressure discharges with magnetic fields,” Ph.D. thesis} (University of Saskatchewan, Canada, 2006). 

\bibitem{Campanell_POP_19_2012} M.~D. Campanell, A.~V. Khrabrov, and I.~D. Kaganovich,
\newblock {\em Phys. Plasmas} 19, 123513 (2012). 

\bibitem{Sheehan_PRL_111_2013} J.~P. Sheehan, N. Hershkowitz, I.~D. Kaganovich, H. Wang, Y. Raitses, E.~V. Barnat, B.~R. Weatherford, and D. Sydorenko, 
\newblock {\em Phys. Rev. Lett.} 111(7), 075002 (2013).

\bibitem{Campanell_APL_103_2013} M. Campanell and H. Wang,
\newblock {\em Appl. Phys. Lett.} 103(10), 104104 (2013).

\bibitem{Carlsson_PSST_26_2016} J. Carlsson, A. Khrabrov, I. Kaganovich, T. Sommerer, and D. Keating, 
\newblock {\em Plasma Sources Sci. Technol.} 26(1), 014003 (2016).

\bibitem{Birdsall_AH_1991} C.~K. Birdsall, 
\newblock {\em Plasma Physics via Computer Simulation} (Adam Hilger, Bristol, 1991).

\bibitem{Hockney_AH_1988} R. W. Hockney and J. W. Eastwood, 
\newblock {\em Computer Simulation Using Particles} (Adam Hilger, Bristol, 1988).

\bibitem{Shahid_JAP_82_1988} R Shahid and Mark J. Kushner, 
\newblock {\em J. Appl. Phys.} 82(6), 2805 (1997).

\bibitem{Lauro_JPDAP_37_2004} Lauro-Taroni, M. M. Turner, and N. StJ Braithwaite, 
\newblock {\em J. Phys. D: Appl. Phys.} 37(16), 2216 (2004).

\bibitem{Lieberman_NJ_2005}
M.~A. Lieberman and A.~J. Lichtenberg,
\newblock {\em Principles of Plasma Discharges and Materials Processing (Wiley,
  NJ, 2005)}, 2005.


\end{thebibliography}

\end{document}